\documentclass[aps,prd,nofootinbib,twocolumn,floatfix,unsortedaddress]{revtex4-1}

\usepackage{graphicx}
\usepackage{dcolumn}
\usepackage{subfigure}
\usepackage{times,mathptm}
\usepackage{float}
\usepackage{color}
\usepackage{amsmath,amsfonts}
\usepackage{mathptmx}
\usepackage{mathrsfs}
\usepackage{bbm}
\usepackage{bm}
\usepackage{xfrac}

\newcommand{\msbar}{{\overline{MS}}}

\newcommand{\beq}{\begin{eqnarray}}
\newcommand{\eeq}{\end{eqnarray}}

\newcommand{\MS}{\overline{MS}}
\newcommand{\LM}{\L_{\MS}}
\renewcommand{\d}{\delta}
\renewcommand{\l}{\lambda}
\renewcommand{\L}{\Lambda}

\renewcommand{\o}{\omega}
\newcommand{\tr}{\text{Tr}}

\newcommand{\vx}{\vec{x}}
\newcommand{\vy}{\vec{y}}
\newcommand{\vz}{\vec{z}}

\newcommand{\vR}{\vec{R}}
\newcommand{\n}{\nu}
\newcommand{\m}{\mu}
\newcommand{\g}{\gamma}
\renewcommand{\r}{\rho}
\newcommand{\e}{\epsilon}
\newcommand{\s}{\sigma}

\newcommand{\V}{{\cal V}}
\newcommand{\D}{\Delta}
\newcommand{\G}{\Gamma}

\newcommand{\vk}{\vec{k}}
\newcommand{\vp}{\vec{p}}

\newcommand{\mk}{|\vec{k}|}

\newcommand{\N}{{\cal N}}

\newcommand{\M}{{\cal M}}
\renewcommand{\th}{\theta}

\newcommand{\oh}{\frac{1}{2}}
\newcommand{\oq}{\frac{1}{4}}

\newcommand{\non}{\nonumber}

\newcommand{\rf}[1]{(\ref{#1})}
\newcommand{\ra}{\rightarrow}
\newcommand{\pa}{\partial}
\renewcommand{\vec}[1]{\bm #1}
\usepackage{ulem}

\begin{document}
\bibliographystyle{h-physrev5}

\title{The Gribov horizon and the one-loop color-Coulomb potential} 

\author{Maarten Golterman}
%, Jeff Greensite$^{1,2}$, Santiago Peris$^{1,3}$,  and Adam P.~Szczepaniak$^4$}
\affiliation{Physics and Astronomy Dept., San Francisco State
University, San Francisco, CA~94132, USA}
\author{Jeff Greensite}
\affiliation{Niels Bohr International Academy, Blegdamsvej 17, DK-2100
Copenhagen \O, Denmark}
\altaffiliation[Permanent address: ]{Physics and Astronomy Dept., San Francisco State
University, San Francisco, CA~94132, USA}
\author{Santiago Peris}
\affiliation{Physics and Astronomy Dept., San Francisco State
University, San Francisco, CA~94132, USA}
\altaffiliation[Permanent address: ]{Departament de F\'{\i}sica, Universitat Aut\`{o}noma de Barcelona, E-08193 Bellaterra, Barcelona, Spain}
\author{ Adam P.~Szczepaniak}
\affiliation{Physics Department and Center for Exploration of Energy and Matter, Indiana University, Bloomington, IN 47403 USA}
% \\
%$^3$ Universitat Aut\`onoma de Barcelona, Barcelona, Spain \\
%$^4$ Physics Department and Center for Exploration of Energy and Matter, Indiana University, Bloomington, IN 47403 USA} 
\date{\today}
\begin{abstract}
 
    We recalculate the color-Coulomb potential to one-loop order, under the assumption that the effect of 
the Gribov horizon is to make i) the transverse gluon propagator less singular; and ii) the color-Coulomb potential more singular, than their perturbative behavior in the low-momentum limit.   As a first guess, the effect of the Gribov horizon is mimicked by introducing a transverse momentum-dependent gluon mass term, leading to a propagator of the Gribov form, with the prescription that the mass parameter should be adjusted 
to the unique value where the infrared behavior of the Coulomb potential is enhanced.   We find that this procedure leads to a Coulomb potential rising asymptotically as a linear term modified by a logarithm.  
  
\end{abstract}

\pacs{11.15.Ha, 12.38.Aw}
\keywords{Confinement,lattice
  gauge theories}
\maketitle

\section{\label{sec:intro}Introduction}

    One of the early ideas regarding the confinement problem was that the confining force might come from one-gluon exchange \cite{Richardson:1978bt,Mandelstam:1979xd,BarGadda:1979cz}.  The suggestion was that a dressed gluon propagator, possibly combined with dressed quark-gluon vertices and arranged in ladder diagrams, would lead to a linear potential.  Of course, the notion that the confining force can be entirely explained by ladder diagrams built from one-gluon exchange must nowadays be considered a little naive.   There are many (related) problems with such a proposal, in particular (i) the existence of a long-range color dipole field around static sources; (ii) long-range van der Waals forces which would then have to exist among hadrons; (iii) group representation dependence (Casimir scaling) rather than N-ality dependence of the asymptotic string tension; and (iv) the absence of color-electric flux tubes, not to mention the absence of string-like properties of such flux tubes, which have been convincingly seen in numerical simulations (cf.\ ref.\ \cite{Athenodorou:2011rx} and references therein). Nevertheless, if it {\it were} possible to reliably calculate the long-range behavior of, say, the color-Coulomb potential, then this information might be useful as an input into more sophisticated pictures, such as the gluon-chain model \cite{Greensite:2001nx}, where the problems just mentioned can be alleviated.   Furthermore, the simple fact is that the instantaneous color-Coulomb potential,  {\it is} linearly confining.  There is ample numerical evidence of this behavior \cite{Greensite:2003xf,Nakagawa:2006fk,Voigt:2008rr}.\footnote{It can be proven that the instantaneous color Coulomb potential is actually an upper bound to the static quark potential \cite{Zwanziger:2002sh}, so even without numerical simulations we would know that the color Coulomb potential must be confining, albeit not necessarily linear.}  

   This article is an an attempt to derive the long-range color Coulomb potential analytically in Coulomb gauge.  There have been a great many efforts in this direction over the years; refs.\ \cite{Zwanziger:2002sh,Alkofer:2009dm,Zwanziger:2003de,Cucchieri:1996ja,Weber:2011zz,Popovici:2010mb,Epple:2006hv,Feuchter:2004mk,Szczepaniak:2001rg} is a partial list, see also refs.\ \cite{Gracey:2009mj,vonSmekal:1997vx,Fischer:2006ub,Alkofer:2008tt} for similar efforts in covariant gauges.  Here we will focus on a simple one-loop perturbative calculation, modified minimally by certain features associated with the Gribov horizon.

\section{Gluon propagators and the Gribov horizon}
 
     The potential energy of two static quarks in color representation $r$ is given in terms of the logarithm of a Wilson loop around a rectangular $R\times T$ contour
\beq
       V_r(R) = - \lim_{T\ra \infty} {1\over T} \log W_r(R ,T)  \ ,
\eeq
where $W_r(R ,T)$ is the vacuum expectation value of the Wilson loop.  Let the side of length $T$ be oriented in the
time direction.  For the purposes of the present article, the color Coulomb potential is defined by counting only the one-gluon exchange contribution to $\log W$, and this is
\beq
   V_C(R) &=& {C_r \over N} \int {d^3 k\over (2\pi)^3}  g^2 N D_{44}(\vk,k_4=0) (1 - e^{i \vk \cdot \vec{R}}) \ ,
\label{VC}
\eeq
where $D^{ab}_{44}(k) = \d^{ab} D_{44}(k)$ is the 44-component of the gluon propagator in Coulomb gauge, $C_r$ is the
quadratic Casimir in representation $r$, and $N$ is the number of colors.   The $R$-independent part of this expression is the self-energy contribution, which we will return to.  It was shown by Zwanziger \cite{Zwanziger:1998ez} that 
$g^2 D_{44}(k)$ is a renormalization group invariant, and therefore does not depend, e.g.\ in the context of dimensional regularization, on the arbitrary scale $\mu$.  $V_C(R)$ is the static quark potential which would be obtained if we approximate the logarithm of a timelike Wilson loop expectation value by the one dressed-gluon exchange term; cf.\  \cite{Susskind:1976pi,Appelquist:1977es,Fischler:1977yf}.    Other treatments focus exclusively on the instantaneous part of $D_{44}$, obtained in the $k_4 \ra \infty$ limit, but in this article we will include also non-instantaneous contributions to the potential, and this leads to setting $k_4=0$. 

    Let us define the renormalization-group invariant
\beq
          V(\vk) \equiv - g^2 N D_{44}(\vk,k_4=0)  \ .
\eeq    
This quantity was computed to one loop long ago \cite{Duncan:1975kt,Appelquist:1977tw}, and the answer (at large $|\vk|$) is
\beq
             V(\vk) = -{1\over \vk^2}{g^2(\mu) N \over 1 + g^2(\m)N {11\over 48 \pi^2} \log{\vk^2 \over \m^2} } \; .
\eeq
Applying the one-loop result 
\beq
             g^2(\m)N = {1\over {11\over 48\pi^2} \log{\m^2 \over \L_{QCD}^2} }
\eeq
we obtain    
\beq
V(\vk) = -{1\over \vk^2}{1 \over {11\over 48 \pi^2} \log{\vk^2 \over \L_{QCD}^2} }   \; ,
\label{Vpt}
\eeq
which is indeed independent, to this one-loop order, of the scale $\m$ introduced in dimensional regularization. 

   However, the perturbative expansion is based on an implicit assumption that, apart from the gauge-fixing condition, the integration over gauge fields is unrestricted; there is no cutoff, for example, in the amplitude of gauge field configurations contributing to the functional integral.  But we have known for many years that this assumption is wrong.  In the lattice formulation, in particular, it is known that
if all gauge copies are included, then the sum over the Faddeev-Popov determinants of each copy will vanish.  This means that the expectation value of any gauge-invariant observable would take on the nonsensical value $0/0$, as was first pointed out by Neuberger \cite{Neuberger:1986xz}.  In the continuum it is also believed, since the seminal work of
Gribov \cite{Gribov:1977wm}, that the functional integral should be restricted to a single gauge copy per gauge orbit,
as in the proposed restriction to the fundamental modular region advocated by Zwanziger \cite{Zwanziger:1998ez}.  It seems difficult to 
implement such a restriction in practice.  At a minimum we can ask that the functional integral be limited to the
Gribov region, in which the lowest eigenvalue of the Faddeev-Popov operator is positive semi-definite, and in fact this is achieved automatically by the gauge-fixing algorithms employed in lattice Monte-Carlo simulations, which find local minima of
\beq
              -\sum_x \sum_{k=1}^3 \tr[U_k(x)] \; .
\eeq
It is the fact that the gauge-fixed configurations are local minima, rather than just stationary points, which ensures that all eigenvalues of the Fadeev-Popov operator are positive.   One can even go a little further.  Since the lattice Monte Carlo procedure will {\it never} generate more than one configuration per gauge orbit in the course of a finite simulation,
an additional restriction to one configuration per orbit is, in some sense, superfluous.\footnote{Of course, if one is interested in a particular selection of gauge copies, such as the fundamental modular region, or the ``B-gauge" \cite{Maas:2009se}, then it is necessary to transform away from the gauge copies generated by the standard algorithms.}

    The limitation to the Gribov region has two expected consequences.  The first, which is true in both Landau and Coulomb gauge, is that the Gribov horizon will impose a cutoff on the magnitude of quantum fluctuations of the transverse gluon
field.  This is easy to check in special cases.  For example, one can construct a (lattice-regularized) plane wave of some fixed amplitude, and compute the low-lying eigenvalues of the lattice Faddeev-Popov operator.  As the amplitude is 
increased, the lowest non-trivial eigenvalue $\l_0$ decreases, and eventually becomes negative.  Configurations with amplitudes such that $\l_0 < 0$ are to be excluded from the functional integration.  

   Gribov \cite{Gribov:1977wm}  suggested that the restriction to the Gribov region would result (in Landau gauge) 
in a gluon propagator of the form
\beq
          D^{ab}_{\m \n}(k) =  \d^{ab} \left( \d_{\m\n} - {k_\m k_\n \over k^2} \right) { 1 \over k^2 + {m^4 \over k^2}} \ ,
\label{GZ}
\eeq
and this propagator clearly vanishes at $k^2 \ra 0$.  Zwanziger \cite{Zwanziger:1988jt}
derived this form by adding to the action
a term which was intended to implement the restriction to the Gribov region.  Gracey \cite{Gracey:2009mj} has calculated the resulting static quark potential to one loop, in Landau gauge, which results from the Zwanziger action.  This potential  turns out to be non-confining.

Lattice simulations, however, have rather
decisively shown \cite{Cucchieri:2010xr,Cucchieri:2007md,Bogolubsky:2009dc} that the Landau gauge gluon propagator has a finite non-zero limit at $k^2 \ra 0$, as is the case for a massive propagator, i.e.
\beq
          D^{ab}_{\m \n}(k) =  \d^{ab} \left( \d_{\m\n} - {k_\m k_\n \over k^2} \right) { 1 \over k^2 + m^2} \ .
\label{CM}
\eeq
   Of course this form cannot be exactly right either; the gluon propagator cannot have a physical pole and must somewhere violate positivity.  Various more complicated forms for the gluon propagator, which agree with \rf{CM} at low momenta, have been put forward, e.g.\ \cite{Aguilar:2011ux}, \cite{Dudal:2008sp}.\footnote{Recently, Zwanziger has suggested a reason why the original proposal in ref.\ \cite{Zwanziger:1988jt} might have failed, c.f.\ ref.\ \cite{Zwanziger:2011yy}.  Dudal et al.\ \cite{Dudal:2008sp} have proposed
a modification of the original Zwanziger action, to bring the result more in line with the lattice results.}   
   
     The corresponding situation in Coulomb gauge is not so clear, at present.  For the transverse gluon propagator at equal times, the Gribov-Zwanziger proposal is that
\beq
          D^{ab}_{ij}(\vk) =  \d^{ab} \left( \d_{ij} - {k_i k_j \over \vk^2} \right) {1 \over 2 \sqrt{\vk^2 + {m^4 \over \vk^2}}} \ ,
\label{GZ1}
\eeq
and numerical calculations by the T\"ubingen group \cite{Burgio:2008jr} seem to support this proposal.  However, other recent calculations by Nakagawa et al.\ \cite{Nakagawa:2011ar} on time-asymmetric lattices, while supporting a vanishing gluon propagator at $k^2\ra 0$, suggest a slower approach to zero than the Gribov-Zwanziger form.  Nakagawa et al.\ conclude that larger lattices will be needed to settle the precise power falloff as $k\ra 0$.   In the absence of decisive lattice data on this point, we will here investigate the consequences of the Gribov-Zwanziger form \rf{GZ1} and also,
for the purpose of contrast, a simple massive transverse propagator
\beq
          D^{ab}_{ij}(k) =  \d^{ab} \left( \d_{ij} - {k_i k_j \over \vk^2} \right) { 1 \over k^2 + m^2} \ .
\label{gprop}
\eeq
Either form is obtained by the naive replacement, in the integration over transverse gauge fields,
\beq
  \lefteqn{\int_G D A_i^{a,tr} \Longrightarrow}
 \non \\ 
 & & \qquad \int DA_i^{tr} \exp\left[-\int {d^4k \over (2\pi)^4} ~ \oh M^2(\vk) A_i^{a,tr}(k)
             A_i^{a,tr}(-k) \right] \ ,
\non \\
\label{mass_term}
\eeq
where
\beq
   M^2(\vk) = \left\{ \begin{array}{cl}
                      m^4/\vk^2 & \mbox{Gribov propagator} \cr
                      m^2 & \mbox{massive propagator} \end{array} \right.
\label{M2}
\eeq
and where the subscript $G$ on the left functional integral refers to the restriction to the Gribov region, with
$A_i^{a,tr}$  the renormalized transverse gauge field.  The replacement is closely related to Zwanziger's 
suggestion \cite{Zwanziger:1988jt}, formulated in Landau gauge, that the restriction to the Gribov region could be implemented by adding an additional term to the action, and this addition includes a mass term with $M^2(k) = m^4/k^2$.
    
    The second expected effect of the restriction to the Gribov region is special to Coulomb gauge.  Coulomb gauge
is a physical gauge, and it has a Hamiltonian containing a non-local  operator
\beq
               {1\over -\nabla \cdot D} (-\nabla^2) {1\over -\nabla \cdot D} \; ,
\eeq
involving two factors of the inverse Faddeev-Popov operator, which is responsible for the Coulomb potential.  Evaluated for a configuration directly on the Gribov
horizon, where the lowest F-P eigenvalue is zero, this quantity is singular.  As Zwanziger has pointed out 
\cite{Zwanziger:1998ez}, we
may expect that most configurations in the Gribov region are quite close to the horizon, for essentially the same reason that most of the volume of a sphere, in a large number of dimensions, is concentrated in the near vicinity of the surface.
But configurations close to the Gribov horizon ought to have an enhanced density of near-zero eigenvalues, as compared to the spectral density of $-\nabla^2$, and a numerical study of configurations generated by lattice Monte Carlo bears this out \cite{Greensite:2004ur}.\footnote{It is interesting that removal of center vortices removes this enhancement, and pushes a typical configuration away from horizon.}  Thus, another effect of restricting configurations to the Gribov region should be an
enhancement of the color-Coulomb potential in the infrared, assuming (as in the free theory) that the infrared behavior is associated with the low-lying eigenmodes of the F-P operator.  

    Thus we are led to explore the consequences of the following two assumptions: first, that the restriction to the Gribov region can be approximately implemented, as in \rf{mass_term}, by the simple addition of a momentum-dependent mass term, and, second, that the value of the mass parameter must be such that the infrared behavior of the Coulomb potential is enhanced beyond the usual $1/\vk^2$ behavior.\footnote{The prescription here is similar to that in ref.\ \cite{Greensite:2010hn}, where a dimensionful parameter in the gluon propagator was adjusted to the precise point where negative Faddeev-Popov eigenvalues disappear.} 
The way in which this could happen is illustrated by the following over-simplified scenario:
The mass term will regularize the infrared behavior of loop integrals, and one might hope (ignoring integrations over Feynman parameters and so on) that the main effect is something like the replacement of $\log(\vk^2/\Lambda^2)$ by $\log((\vk^2 + m^2)/\Lambda^2)$ in eq.\ \rf{Vpt}.  Then, just by tuning $m=\Lambda$, the color-Coulomb potential
at low momentum becomes
\beq
  V(\vk) &\sim&  -{1 \over \vk^2 \log\left({\vk^2 + \L^2\over \L^2}\right)}
\non \\
             &\sim&  -{\L^2 \over \mk^4} \ ,
\eeq
much as in the old Richardson proposal \cite{Richardson:1978bt}. We will now see how close we can come to
realizing this scenario.

\section{One-Loop Integrals in First-Order Formalism}

   The Coulomb potential is directly related to the 44 component of the gluon propagator.  If we denote by 
$\d^{ab} \Pi_{\m \n}$ the one-particle irreducible contribution to the Coulomb gauge gluon propagator, and noting that 
$\Pi_{4 i}=0$ for $i\ne 4$, then the 44 component can be expanded, as usual, in a geometric series
\beq
      D_{44}(k) &=& {1\over \vk^2} \Bigl( 1 +   
                      \Pi_{44}(k){1\over \vk^2} + \left(\Pi_{44}(k){1\over \vk^2}\right)^2 + ... \Bigr)
\non \\
                      &=& {1\over \vk^2} {1 \over 1 - g^2 N\Pi(k)} \ ,
\eeq
where $g^2 N\Pi(k) \equiv \Pi_{44}(k)/ \vk^2$.   We would then like to calculate 
$\Pi(\vk,k_4=0)$ to one loop, with the restriction to the Gribov region approximated by adding a mass term to the gauge-fixed action.    Even at the one loop level,
the loop integrals are complicated and non-covariant, and some are difficult to evaluate by standard formulas.  It turns out to be much simpler to carry out the calculation in the first-order formulation, which is often used when dealing with Yang-Mills theory quantized in Coulomb gauge (see in particular  
\cite{Zwanziger:1998ez,Feinberg:1977rc,Cucchieri:2000hv,Watson:2007mz}).

    The starting point for the first-order formalism is the Euclidean partition function for 
Yang-Mills theory fixed to Coulomb gauge
\beq         
Z(J) &=& \int_G DA_\m \d[\nabla \cdot A]  \det[\M] 
\non \\
& & \qquad \times \exp\left[-\int d^4x (\oq F^2_{\m \n} + ig J_\m A_\m)\right] \ ,
\eeq
where $\M = - \nabla \cdot D$ is the Faddeev-Popov operator, and the color indices on the gauge field and field strength tensor are not written out explicitly, but are left implicit.
One then introduces an $E_i$ field via the identity.
\beq
\exp\left[-\oh \int d^4x F_{0i}^2\right] = {\cal \N} \int DE_i \exp\left[\int (iE_i F_{0i} - \oh E_i^2) \right] \ .
\non \\
\eeq
The $E$ field is split into a transverse and longitudinal piece $E_i=E_i^{tr} - \pa_i \phi$, and then one integrates
out the $A_4$ field, which generates a delta-function enforcing the Gauss Law constraint.  This is followed by integration
over the $\phi$ field, which eliminates both the Faddeev-Popov determinant and the Gauss Law delta function.  
The details of how this goes can be found, e.g., in ref.\ \cite{Cucchieri:2000hv}, and the result is
\beq
\lefteqn{Z[J] =}
\non \\
& & \int_G DA_i^{tr} \int DE_i^{tr} \exp\left[ \int d^4x \Bigl(iE_i^{tr} \dot{A}_i^{tr} -  \oh (E_i^{tr 2}  + B_i^2) 
-  i g J_i A_i^{tr}\Bigr) \right.
\non \\
& & \left. - \oh \int dt d^3x d^3y (\r_C + gJ_4)_{\vx,t} K[\vx,\vy,t,A^{tr}]  (\r_C + g J_4)_{\vy,t} \right] \ ,
\non \\
\eeq
where
\beq
\r^a_C(x) = -g f^{abc} A_i^{b,tr}(x) E_i^{c,tr}(x) 
\eeq
and $B_i^a = \oh \epsilon_{ijk}F_{jk}$ is constructed from the transverse $A$-field.
The non-local kernel, providing the Coulombic part of the Coulomb-gauge Hamiltonian, is  
\beq
 K[\vx,\vy,t,A^{tr}] = \left[\M^{-1} (-\nabla^2) \M^{-1}\right]^{ab}_{\vx,\vy} \ .
\eeq
Then
\beq
 \d^{ab} D_{44}(x-y) &=&  -\left[ {1\over g^2 Z} {\d^2 \over \d J^a_4(x) \d J^b_4(y)} Z  \right]_{J=0}
\non \\
&=&  \langle K^{ab}(\vx,\vy,A^{tr}(x_4)) \rangle \d(x_4-y_4)
\non \\
& & - \left\langle \int d^3z_1 K^{ac}(\vx,\vz_1,A^{tr}(x_4)) \r^c(\vz_1,x_4) \right.
\non \\
& & \times \left. \int d^3z_2 K^{bd}(\vy,\vz_2,A^{tr}(y_4)) \r^d(\vz_2,y_4) \right\rangle \ .
\non \\
\eeq

    The contribution to one loop is obtained by expanding $K^{ab}(\vx,\vy,A^{tr}(x_4))$ up to second order
in the coupling.  Since the product $\r \r$ inside the integrals over $\vz_1,\vz_2$ is already second order,
we can set $K$ to its zeroth-order value in the integrand.  The result is
\begin{widetext}
\beq
 \lefteqn{\d^{ab} D_{44}(x-y)}
 \non \\
  & & = \d^{ab}\left[ \left({1 \over -\nabla^2} \right)_{\vx,\vy}  
     + 3g^2 f^{acd}f^{dfb} \int d^3z_1 d^3z_2  \left({1 \over -\nabla^2} \right)_{\vx,\vz_1} 
     \langle A_i^c(z_1) A_j^f(z_2)\rangle_0  
      (\pa_i)_{\vz_1} \left({1 \over -\nabla^2} \right)_{\vz_1,\vz_2}(\pa_j)_{\vz_2} \left({1 \over -\nabla^2} \right)_{\vz_2,\vy} \right]  \d(x_4-y_4)
\non \\
& & \qquad -g^2 f^{acd}f^{bef} \int d^3z_1 d^3z_2  \left({1 \over -\nabla^2} \right)_{x,\vz_1}\left\{
\langle A_i^c(\vz_1,x_4) A_j^e(\vz_2,y_4) \rangle_0 \langle E_i^d(\vz_1,x_4) E_j^f(\vz_2,y_4) \rangle_0 \right.
\non \\
& & \left. \qquad \qquad
+ \langle A_i^c(\vz_1,x_4)  E_j^f(\vz_2,y_4) \rangle_0 \langle  A_j^e(\vz_2,y_4) E_i^d(\vz_1,x_4)   \rangle_0  \right\}
\left({1 \over -\nabla^2} \right)_{\vz_2,\vy} \ .
\label{d44}
\eeq

\end{widetext}

   In ordinary perturbation theory, the zeroth-order propagators are determined by simply removing the restriction
to the Gribov region in the integral over $A^{tr}_i$.  Introducing polarization vectors
\beq
      A_{i}^{a,tr}(k) = \sum_{\l=1}^2 \epsilon_i^\l(k) A^a(k,\l) 
\eeq
with the usual properties
\beq
 k_i \epsilon_i^\l(k) = 0 ~~~,~~~\epsilon^{\l*}_i(k) \epsilon^{\l'}_i(k) = \d^{\l \l'}  \ ,
\eeq
and
\beq
          T_{ij} &\equiv& \sum_\l \epsilon^{\l*}_i(k) \epsilon^\l_j(k)
\non \\
                   &=& \d_{ij} - {k_i k_j \over \vk^2} \ ,
\eeq
so that
\beq
\int DA^{a,tr}_i(k) = \int DA^a(k,\l) \ ,
\eeq
one can easily derive the zeroth-order momentum-space propagators in first-order formalism
\beq
  \langle A_i^a(k) A_j^b(k')\rangle_0 &=& \d^{ab} T_{ij}(\vk) {1 \over k^2} \d^4(k+k') \ ,
\non \\
  \langle E_i^a(k) E_j^b(k')\rangle_0 &=& \d^{ab} T_{ij}(\vk) {\vk^2 \over k^2} \d^4(k+k') \ ,
\non \\
  \langle E_i^a(k) A_j^b(k')\rangle_0 &=& \d^{ab} T_{ij}(\vk) {k_4 \over k^2} \d^4(k+k') \ .
\eeq  
Taking eq.\ \rf{d44} to momentum space and inserting the propagators above, one finds for
$\Pi(k)$
\beq
\Pi(k) &=& {1 \over \vk^2}\left\{  3 k_i k_j \int {d^4p \over (2\pi)^4} {T_{ij}(\vp) \over p^2 (\vp-\vk)^2} \right.
\non \\
    & & \left. - \int  {d^4p \over (2\pi)^4} {T_{ij}(\vp) \over p^2}{T_{ij}(\vp-\vk) \over (p-k)^2}[\vp^2 - p_4(p_4-k_4)] \right\} \ ,
\non \\
\eeq
as originally obtained in ref.\ \cite{Cucchieri:2000hv}, see also \cite{Watson:2007mz}.   The integrals can be
evaluated under dimensional regularization, and the standard result for the one-loop momentum space Coulomb potential is obtained.

    Now suppose that instead of simply removing the restriction to the Gribov horizon in the integration over $A^{tr}$, we try to mimic its effect by insertion of a mass term, as in eq.\ \rf{mass_term}.  The effect on the zeroth-order propagators
is readily obtained:
\beq
  \langle A_i^a(k) A_j^b(k')\rangle_0 &=& \d^{ab} T_{ij}(\vk) {1 \over k^2 + M^2(\vk)} \d^4(k+k') \ ,
\non \\
  \langle E_i^a(k) E_j^b(k')\rangle_0 &=& \d^{ab} T_{ij}(\vk) {\vk^2 + M^2(\vk) \over k^2  + M^2(\vk)} \d^4(k+k') \ ,
\non \\
  \langle E_i^a(k) A_j^b(k')\rangle_0 &=& \d^{ab} T_{ij}(\vk) {k_4 \over k^2 + M^2(\vk)} \d^4(k+k') \ .
\eeq
The (unregulated) expression for $\Pi(\vk)$, in the $k_4=0$ case we consider here,  then becomes
\beq
 \Pi(\vk) =  J_1 - J_2 \ ,
\label{pik}
 \eeq
 where
\beq
  J_1 &=&  3 {k_i k_j \over \vk^2} \int {d^4p \over (2\pi)^4} {T_{ij}(\vp) \over (p^2 + M^2(\vk))  (\vp-\vk)^2} 
\non \\  
  &=&  {3\over 2} {k_i k_j \over \vk^2}  \int {d^3p \over (2\pi)^3} 
  {T_{ij}(\vp) \over (\vp^2+M^2(\vk))^{1/2}  (\vp-\vk)^2} \ ,
\label{J1}
\eeq
and
\beq
    J_2 &=&   \int  {d^4p \over (2\pi)^4} {T_{ij}(\vp) \over p^2 + M^2(\vk)}{T_{ij}(\vp-\vk) \over (p-k)^2 + M^2(\vk)}
\non \\
     & & \qquad \qquad  \times [\vp^2  + M^2(\vp) - p_4^2]  
\non \\
&=& {1\over 2 k^2} \int  {d^3p \over (2\pi)^3} {\o_p - \o_{p-k} \over \o_p + \o_{p-k}} {T_{ij}(\vp) 
T_{ij}(\vp-\vk) \over \o_{p-k}} \ ,
\label{J2}
\eeq
with 
\beq
          \o_p \equiv \sqrt{\vp^2 + M^2(\vp)} \ .
\label{omega}
\eeq    
Our task is to evaluate suitably regularized versions of $J_{1,2}$ for the two choices of $M^2(\vk)$ shown in 
eq.\ \rf{M2}. \\

\section{Dimensional regularization, massive propagator \label{massive_case}}

    As a first step, we will compute the Coulomb potential to one loop using the massive transverse gluon propagator shown in eq.\ \rf{gprop}.  We do not believe this propagator is correct in Coulomb gauge even at low momenta.  In contrast to Landau gauge, existing lattice simulations indicate an equal-times propagator which falls to zero at $\vk^2=0$, as already mentioned.  
The massive propagator is mainly useful as an illustration of how the potential can be enhanced by appropriately tuning the mass parameter, and also serves as a contrast to the results obtained in the next section.   Technically, the massive propagator is simpler than the Gribov propagator case, in that standard dimensional regularization can be applied without 
any difficulty to the relevant loop integrals.
    
    We now apply dimensional regularization, taking into account
the fact that $\d_{ii}=3-2\e$.  Then 
\beq
\Pi(\vk) &=&   {3\over 2}\left(I_1 -  I_2 \right)
                       - {2-2\e \over \vk^2}(I_{3a} + I_{3b} - I_4) + {1\over \vk^2}I_5   \  ,
\non \\
\label{Pik}
\eeq
where
\beq
I_1 &=& \m^{2\e} \int {d^{2\o'}p \over (2\pi)^{2\o'}} {1 \over (\vp^2 + m^2)^{1/2} (\vp - \vk)^2 } \ ,
\non \\
I_2 &=& {k_i k_j \over \vk^2} \m^{2\e} \int {d^{2\o'}p \over (2\pi)^{2\o'}}  {p_i p_j \over \vp^2 (\vp^2 + m^2)^{1/2} (\vp - \vk)^2 } \ ,
\non \\
I_{3a} &=&\m^{2\e} \int {d^{2\o}p \over (2\pi)^{2\o}} {m^2 \over (p^2 + m^2)((p-k)^2 + m^2)} \ ,
\non \\
I_{3b} &=& \m^{2\e}\int {d^{2\o}p \over (2\pi)^{2\o}} {\vp^2 \over (p^2 + m^2)((p-k)^2 + m^2)} \ ,
\non \\
I_{4} &=& \m^{2\e}\int {d^{2\o}p \over (2\pi)^{2\o}} {p_4^2 \over (p^2 + m^2)((p-k)^2 + m^2)} \ ,
\non \\
I_5 &=& \m^{2\e}\int {d^{2\o}p \over (2\pi)^{2\o}} {\vp^2 + m^2 - p_4^2 \over (p^2 + m^2)((p-k)^2 + m^2)} \ ,
\non \\
   & & \qquad  \times {\vp^2 \vk^2 - (\vp \cdot \vk)^2 \over \vp^2 (\vp-\vk)^2} \ ,
\eeq
and $\o' = {3\over 2} -\e, ~ \o = 2-\e$.  Integrals $I_1$ through $I_4$ are divergent, $I_5$ turns out to be finite.
Before carrying out the usual $\overline{MS}$ subtractions, it is important to note that one is only allowed to make the subtractions which are made at $m^2=0$.  In particular, one cannot subtract terms proportional to $m^2$, because there is no counterterm which would generate such a subtraction. 

The integrals can all be evaluated by the standard methods, and the results for the divergent integrals are
\begin{widetext}
\beq
I_1 &=& {1\over 4\pi^2} \left({1\over \e} - \g + \log 4\pi \right) -{1\over 8\pi^2} \int  dx ~ x^{-1/2} \log\left({\vk^2 x(1-x) + m^2 x \over \mu^2}\right) \ ,
\non \\
I_2 &=&  {1\over 12\pi^2} \left({1\over \e} - \g + \log 4\pi \right) - {1\over 16\pi^2}  \int_0^1 dx_1 dx_2 ~ \theta(1-x_1-x_2) x_1^{-1/2} 
           \log\left({\vk^2 x_2(1-x_2) + m^2 x_1 \over \mu^2}\right)  
\non \\
   & & + {\vk^2 \over 8\pi^2} 
           \int  dx_1 dx_2 ~ \theta(1-x_1-x_2) {x_1^{-1/2} x_2^2 \over \vk^2 x_2(1-x_2) + m^2 x_1}  \ ,
\non \\
I_{3a} &=& {m^2 \over (4\pi)^2}\left\{ \left({1\over \e} - \g + \log 4\pi \right) - \int  dx ~ \log\left({\D \over \m^2}\right) 
\right\} \ ,
\non \\
I_{3b} &=& {1\over 48 \pi^2} \vk^2 \left({1\over \e} - \g + \log 4\pi \right) 
- {\vk^2 \over (4\pi)^2} \int  dx ~x^2 \log\left( {\D \over \m^2} \right) 
- {3\over 2} {1 \over (4\pi)^2} \left({1\over \e} - \g + \log 4\pi + {1\over 3}  \right)\left({\vk^2 \over 6} + m^2\right)
\non \\ 
& & + {3\over 2}{1\over (4\pi)^2} \int  dx ~ \D \log\left( {\D \over \m^2} \right)  \ ,
\non \\
I_4 &=&  -\oh {1 \over (4\pi)^2}  \left({1\over \e} - \g + 1 + \log 4\pi \right)\left({\vk^2 \over 6} + m^2\right)  +
 {1\over 2}{1 \over (4\pi)^2} \int  dx ~  \D \log\left( {\D \over \m^2} \right)   \ .
\eeq
\end{widetext}
In these expressions we have defined 
\beq
\D \equiv \vk^2 x(1-x) + m^2 \  .
\eeq
All $x$-integrations run from 0 to 1, and $\th(x)$ is the Heaviside theta function.
 
At this point we should take note of a source of possible trouble.  In the first place, some of the integrals have produced
$m^2 / \e$ terms, which cannot be subtracted away. 
Even finite terms proportional to $m^2$ would be catastrophic to our program, because these would tend to make
the color Coulomb potential less, rather than more, divergent in the infrared.  Somewhat remarkably, when the above 
integrals are inserted into \rf{Pik}, we find that there is a complete cancellation of the dangerous terms proportional to $m^2$, while the remaining terms proportional to $1/\e - \g + \log 4\pi$ can be subtracted in the usual way.  The end result is that
\beq
\lefteqn{\Pi(\vk,\m) = }
\non \\
 & & -{3\over 16 \pi^2} \int dx ~ x^{-1/2} \log\left({k^2 x(1-x) + m^2 x \over \mu^2}\right) 
\non \\
& & + {3\over 32 \pi^2} \int dx_1 dx_2 \theta(1-x_1-x_2) x_1^{-1/2} 
\non \\
& &  \qquad\times
           \log\left({\vk^2 x_2(1-x_2) + m^2 x_1 \over \mu^2}\right) 
\non \\
& &  - {3\over 16 \pi^2} \vk^2 \int dx_1 dx_2 \theta(1-x_1-x_2) {x_1^{-1/2} x_2^2 \over \vk^2 x_2(1-x_2) + m^2 x_1} 
\non \\
& & - {1\over 8\pi^2}\int dx ~ x(1-2x) \log\left({\D \over \mu^2}\right)  + {1\over 48\pi^2} + {1\over \vk^2} I_5   \ ,
\eeq   
and therefore
\beq
V(\vk) = -g^2(\m) N D_{44} = -{1\over \vk^2} {1\over {1\over g^2(\m) N} - \Pi(\vk,\m)} \ .
\eeq
Inserting the one-loop expression for $g^2(\m)$, one finds that the dimensional regularization scale $\m$ cancels out
exactly, leaving the result
\beq
V(\vk) = {1 \over \vk^2 \Pi(\vk,\LM)} \ .
\eeq

    Now we consider the infrared limit, $\vk^2/m^2 \ll 1$, starting with the integral $I_5$. Although this integral looks
superficially divergent, it is clear, after an integration over $p_4$ which gives
\beq
I_5 &=& \oh \int {d^3 p \over (2\pi)^3}  {\vp^2 \vk^2 - (\vp \cdot \vk)^2 \over \vp^2 (\vp-\vk)^2}
{1\over \sqrt{(\vp-\vk)^2 + m^2}}
\non \\
& & \qquad \times { \sqrt{\vp^2 + m^2} - \sqrt{(\vp-\vk)^2 + m^2} \over 
\sqrt{\vp^2 + m^2} + \sqrt{(\vp-\vk)^2 + m^2} } \ ,
\eeq
that in fact the integral is finite.  Although it is still complicated, it is not hard to show that the low-momentum limit,
up to $O(k^2/m^2)$, is rather simple:
\beq
{1\over \vk^2} I_5 = {1\over 360 \pi^2} {\vk^2 \over m^2} \ .
\eeq
It is also simple to evaluate the low-momentum limit of the single integrations over $x$:
\beq
\lefteqn{\int dx ~ x^{-1/2} \log\left({\vk^2 x(1-x) + m^2 x \over \LM^2}\right)}
\non \\
& & \qquad \qquad \qquad \ra {4\over 3}{\vk^2 \over m^2} + 2 \log{m^2 \over \LM^2} - 4 \ ,
\non \\
\lefteqn{\int dx ~ x(1-2x) \log\left({\vk^2 x(1-x) + m^2 \over \LM^2}\right)}
\non \\
& & \qquad \qquad \qquad \ra -{1\over 60} {\vk^2 \over m^2} - {1\over 6} \log{m^2 \over \LM^2} \ .
\non \\
\eeq
If this were all there were, then it would be possible to choose $m \propto \LM$ so as to cancel the constant terms,
leaving only a term proportional to $\vk^2/m^2$.  This would lead to an overall $1/\vk^4$ dependence for the
color Coulomb potential, and therefore to a linear potential.  However, the
integral $I_2$ leads to the two expressions involving integration over two Feynman parameters, and these turn
out to spoil the desired result.  The double integrals can be evaluated analytically at low $k^2$, with the help  of the Mellin-Barnes transform and converse mapping theorem \cite{Flajolet19953,Friot:2005cu}.  The details are reserved for Appendix A. The result, up to $O(k^2/m^2)$, is
\beq
& & \int dx_1 dx_2 \theta(1-x_1-x_2) x_1^{-1/2} 
           \log\left({\vk^2 x_2(1-x_2) + m^2 x_1 \over \LM^2}\right) 
\non \\
& & \qquad = {4\over 3} \log{m^2 \over \LM^2} - {32 \over 9} 
+ {\pi^2 \over 4} \left({\vk^2 \over m^2}\right)^{1/2} - {8\over 15} {\vk^2 \over m^2} \ ,
\label{bad1}
\eeq
and
\beq
& & \vk^2 \int dx_1 dx_2 \theta(1-x_1-x_2) {x_1^{-1/2} x_2^2 \over \vk^2 x_2(1-x_2) + m^2 x_1} 
\non \\ 
& & \qquad = {3 \pi^2 \over 8}  \left({\vk^2 \over m^2}\right)^{1/2} 
 - {32\over 15} {\vk^2 \over m^2} \ .
\label{bad2}
\eeq
Note the appearance of terms proportional to $|\vk|$.
Therefore, at low momenta,
\beq
V(\vk) = -{1\over \vk^2} {\pi^2 \over {11\over 48}\log{m^2 \over \LM^2} - {7\over 16} +  {3 \pi^2 \over 64} \left({\vk^2 \over m^2}\right)^{1/2} - {151 \over 1440}{\vk^2 \over m^2}} \ . \non \\
\label{Vmassive}
\eeq
We have suggested that $m$ should be set to the unique value which would enhance the infrared behavior of the Coulomb
potential.  This value is 
\beq
            m = e^{21/22} \LM \ ,
\eeq
leading to the final result at low momentum: 
\beq
V(\vk) = - {64\over 3} e^{21/22} {\LM \over |\vk|^3} \ .
\eeq
Since the term proportional to $\mk^3$ is dominant at low momenta, this results in an asymptotic potential rising 
logarithmically with quark separation.

\section{Cutoff regulator, Gribov propagator}

    The result found in the previous section would be a little disappointing, if the transverse gluon propagator actually had the massive form with $M^2(\vk)=m^2$.
Tuning the mass parameter to the unique value which enhances the Coulomb potential does take us to a potential which rises faster than $1/r$, but the rise is still only logarithmic at large color charge separation.   We will now investigate what happens in the (possibly) more realistic case where the transverse gluon propagator takes on the Gribov form.

   We again have $\Pi(k) =  J_1(k) - J_2(k)$, where $J_{1,2}$ are given in eqs.\ (\ref{J1}-\ref{omega}), but this time
with the choice $M^2(\vp) = m^4/\vp^2$.  It is awkward to evaluate $J_2$, in particular, by dimensional regularization; one would end up with a complicated multiple integral over very many Feynman parameters.  Since we are only interested in the small-$k^2$ behavior of these integrals, we have found it convenient to follow a different strategy, based on a simple momentum cutoff at $|\vp|=\L$.  

   We are aware that a momentum-cutoff regulator is dangerous in gauge theories, and is likely to violate Ward identities and introduce spurious divergences, but these problems will not arise in our present one-loop calculation.
This does not mean that the momentum cutoff procedure is necessarily consistent at higher loops, but that property is not crucial to us.  What we are really after is to use the momentum cutoff result to figure out what the one-loop result for $V(k)$ would be in the $\MS$ scheme, without actually evaluating the integrals via dimensional regularization.  This strategy requires that the momentum cutoff and dimensional regularization results can be matched exactly at one loop, by an appropriate choice of coupling $g^2$ in the cutoff regularization.   That matching will be postponed to the next section.

   From this point on, since we will mainly be carrying out integration in three dimensions, we will denote 
\beq
           k = |\vk| ~~,~~ p = |\vp| \, .
\eeq
Of course the first equality is true even if $k$ denotes the modulus of the 4-momentum, since we only consider the case 
where $k_4=0$.

   Begin with $J_1$, which, with a momentum cutoff, can be written as
\beq
    J_1(k) &=& {3\over 2} {1\over 4 \pi^2} \int_0^\L dp p^2 \int_{-1}^1 du {p \over \sqrt{p^4 + m^4}}{ 1 - u^2 \over p^2 + k^2 - 2pku} \ ,
\non \\
\eeq
and make the split
\beq
{1 \over \sqrt{p^4 + m^4}} = {1\over m^2} + {m^2 - \sqrt{p^4 + m^4} \over m^2 \sqrt{p^4 + m^4}} \ ,
\eeq
so that
\beq
J_1(k) &=& {3\over 2}(J_{1A} + J_{1B})
\eeq
where
\beq
J_{1A} &=&  {1\over 4 \pi^2} \int_0^\L dp p^3 \int_{-1}^1 du {1-u^2 \over m^2 (p^2 + k^2 - 2pku)} \ ,
\non \\
J_{1B} &=&  {1\over 4 \pi^2} \int_0^\L dp p^3 \int_{-1}^1 du {m^2 - \sqrt{p^4 + m^4} \over m^2 \sqrt{p^4 + m^4}}
\non \\
 & & \qquad \times {1-u^2 \over p^2 + k^2 - 2pku} \ .
\eeq
Integral $J_{1A}$ can be evaluated analytically, with the result up to $O(k^2/m^2)$ (and discarding terms of
$O(1/\L^2)$)
\beq
J_{1A}
  = \frac{\L^2}{6 \pi ^2 m^2} +{k^2 \over m^2} \frac{  \left(15 \log \left({k^2 \over \L^2}\right) -46\right)}{450 \pi ^2}  \ .
\eeq

    For the integral $J_{1B}$ we first expand in powers of $k^2$ the term
\beq
 \lefteqn{{1-u^2 \over p^2 + k^2 - 2pku}}
\non \\ &=& \frac{1-u^2}{p^2} -\frac{2 k
   \left(u^3-u\right)}{p^3}
 +\frac{k^2 \left(-4 u^4+5 u^2-1\right)}{p^4} + O(k^3) \ , \non \\
\eeq
and find, again up to $O(k^2/m^2)$ and discarding terms of $O(1/\L^2)$,
 \beq
J_{1B} &=& {1 \over 60 \pi ^2 m^2}
\left[ 2 k^2 \log {\L^2 \over 2m^2}  -10 \L^2+10 m^2 \log  {2\L^2 \over m^2}   \right] \ .
\non \\
\eeq
Adding together $J_{1A}$ and $J_{1B}$, we then have
\beq
J_1 &=& {1\over 300 \pi^2}\left\{ 15 {k^2\over m^2} \left(\log \frac{k^2}{m^2}-\log 2 \right)
   -46 {k^2\over m^2} \right.
\non \\
& & \left. \qquad +75 \left(\log\frac{\L^2}{m^2}+\log 2\right) \right\}  \ .
\eeq   
Both $J_{1A}$ and $J_{1B}$ are quadratically divergent, but this is only an artifact of splitting $J_1$ into two pieces.  These quadratic divergences cancel exactly in the sum, as they must, since the $J_1$ integral is only logarithmically divergent in the cutoff $\L$.

   We employ a similar strategy to evaluate $J_2(k)$ at low momenta. Defining $R_p = \sqrt{p^4 + m^4}$, 
the integrand in \rf{J2} is
\beq
F_1(p,k,u) &=&  {|\vp-\vk| \over R_{p-k}}  {|\vp-\vk| R_p - p R_{p-k} \over  |\vp-\vk|R_p + p R_{p-k} }
\non \\
    & & \; \; \times    \left( {1\over k^2} - \oh {1-u^2 \over p^2 + k^2 - 2pku} \right) \ .
\eeq
Let $F_0(p,k,u)$ be the same expression with $R_p, R_{p-k}$ both replaced by $m^2$
\beq
F_0(p,k,u) =  {|\vp-\vk| \over m^2} {|\vp-\vk|   - p  \over  |\vp-\vk|  + p   }
                         \left( {1\over k^2} - \oh {1-u^2 \over p^2 + k^2 - 2pku} \right) \ .
\non \\
\eeq 
Then we write $J_2 = J_{2A} + J_{2B}$ where
\beq
J_{2A}(k) &=&  {1\over 4\pi^2}  \int_0^\L dp p^2 \int_{-1}^1 du F_0(p,k,u) \ ,
\non \\
J_{2B}(k) &=&  {1\over 4\pi^2}  \int_0^\L dp p^2 \int_{-1}^1 du (F_1(p,k,u)-F_0(p,k,u)) \ . \non \\
\eeq
The first integral can be done analytically, and again keeping terms to $O(k^2/m^2)$ and dropping $O(1/\L^2)$,
\beq
J_{2A}(k) = \frac{\L^2}{16 \pi ^2 m^2}+\frac{k^2 (105 \log {k\over \L} -345 \log 2
      +83)}{7200 \pi ^2 m^2}  \ .
\non \\
\eeq
 
  To evaluate $J_{2B}$, we expand the integrand in a power series in $k$.  Then the integration over $p$
and $u$ can be carried out, with the result
\beq
J_{2B} &=& {7 \over 960 \pi^2}\frac{k^2}{m^2} \log {\L^2 \over 2m^2} +\frac{127}{7200
   \pi ^2}{k^2 \over m^2} 
   \non \\   
   & & - {1\over 16 \pi ^2}{\L^2 \over m^2} + {1\over 48\pi^2}\log{2 \L^2 \over m^2} -\frac{5}{72 \pi ^2} \ .
\eeq
Combining $J_{2A}$ and $J_{2B}$
\beq
J_2 &=&   {k^2\over m^2}  \left({7\over 960 \pi^2} \log {k^2 \over m^2}
   +\frac{7}{240 \pi ^2}-\frac{53 \log 2}{960 \pi
   ^2}\right) 
\non \\
& &   + {1\over 48\pi^2} \log{2\L^2 \over m^2} 
   -\frac{5}{72 \pi ^2}    \ .
\eeq
As with $J_1$, the quadratically divergent terms in $J_{2A}$ and $J_{2B}$ necessarily cancel in the sum, since the $J_2$ integral is only logarithmically divergent in $\L$.

   Substituting the results for $J_1$ and $J_2$ into \rf{pik}, we now have, up to O$(k^2)$,
\begin{widetext}
\beq
\Pi(k) =  {k^2\over m^2} \left( {41 \over 960 \pi^2} \log{k^2 \over m^2} - {73 \over 400 \pi^2} + 
{\log 2 \over 192 \pi^2} \right)
  +{11\over 48 \pi^2} \log \frac{2\L^2}{m^2}+\frac{5}{72 \pi ^2}   \ ,
\eeq
and therefore
\beq
V(k) = - {1\over k^2}\Bigg[ {1\over g^2 N} -   {k^2\over m^2} \left( {41 \over 960 \pi^2} \log{k^2 \over m^2} - {73 \over 400 \pi^2} + 
{\log 2 \over 192 \pi^2} \right)
  -{11\over 48 \pi^2} \log \frac{2\L^2}{m^2}-\frac{5}{72 \pi ^2}\Bigg]^{-1}  \ . \non \\
\label{Vinv}
\eeq
\end{widetext}
%Note the factor of $11 / 48 \pi^2$ in front of the logarithmically divergent term.  This will be important in what follows.

\section{Connecting the regulators}

    Now that we have computed $V(k)$ to one loop with a momentum cutoff, the task is to figure out what the result would
have to be in the $\MS$ scheme, because we would like to express our result in terms of a physical scale 
such as $\LM$.  The key is to show that it is possible to choose $g^2=g^2(\L)$ in the cutoff expression, such that an exact matching to $\MS$ is possible.  

     Denote
\beq
    \Pi(k) = \int [dp] R(p,k,m) \ ,
\eeq
where $[dp]$ denotes the multiple integration measure, and of course the integral is logarithmically divergent.
What we would like to calculate is
\beq
    [-k^2 V(k)]^{-1} =   {1\over g^2_\msbar(\m)N} - \int_\msbar [dp] R(p,k,m) \ ,
\eeq
where the integral is dimensionally regulated, and the usual $\msbar$ subtractions are carried out.  What we actually compute, however, is 
\beq
   {1\over g^2(\L)N} -  \int_\L [dp] R(p,k,m) \ ,
\eeq
where the integral is regulated with a momentum cutoff, and the dependence of the coupling on the cutoff is not yet specified.  This expression can be rewritten slightly as
\beq
  {1\over g^2(\L)N} &-& \int_\L [dp] R(p,k,0) 
\non \\
  &-&  \int [dp] \{ R(p,k,m)-R(p,k,0) \} \ ,
\eeq
where the second integration is finite, and needs no regulator.   Now suppose it is possible to choose $g^2(\L)N$
such that, as $\L \ra \infty$
\beq
 \lefteqn{{1\over g^2(\L)N} - \int_\L [dp] R(p,k,0)}
 \non \\ 
 & & \qquad \qquad  = {1\over g^2_\msbar(\m)N} - \int_\msbar [dp] R(p,k,0) \ .
\label{matching}
\eeq
Then
\beq
\lefteqn{{1\over g^2(\L)N} - \int_\L [dp] R(p,k,m)}
\non \\
& & \qquad = {1\over g^2(\L)N} - \int_\L [dp] R(p,k,0) 
\non \\
& & \qquad \qquad -  \int [dp] \{ R(p,k,m)-R(p,k,0) \}
\non \\
& & \qquad = {1\over g^2_\msbar(\m)N} - \int_\msbar [dp] R(p,k,0)
\non \\
& & \qquad \qquad -  \int [dp] \{ R(p,k,m)-R(p,k,0) \}
\non \\
& & \qquad = {1\over g^2_\msbar(\m)N} - \int_\msbar [dp] R(p,k,m) \ .
\eeq
The conclusion is that if we can find a  $g^2(\L)N$ which satisfies the matching condition \rf{matching} at $m^2=0$, then the cutoff-regulated calculation will give us the desired result in the $\msbar$-scheme for any $m^2$.

%   It is important to note that, whatever the regulator, the dependence of the integrals on $m^2$ is the same.  To see this,
%let $F(p,k,m)$ be the integrand of one of the logarithmically divergent integrals we are concerned with, and let
%$\cal{R}$ denote the regulator.  Unless the regulator is truly pathological, it must be true that
%\beq
%   \lefteqn{\int_{\cal{R}} {d^3 p \over (2\pi)^3} F(p,k,m) = \int_{\cal{R}} {d^3 p \over (2\pi)^3} F(p,k,0)}
%\non \\
%        & & \qquad \qquad + \int {d^3 p \over (2\pi)^3} [F(p,k,m)-F(p,k,0)]
%\eeq
%Since the last integral on the right is finite and therefore regulator independent, if follows that the $m^2$ dependence of the integral on the left is regulator independent.  So $m^2$ dependence in the cutoff and dimensional regularizations of $J_1-J_2$ should match automatically, without any appeal to $g^2(\L)$.  To determine $g^2(\L)$, it is therefore sufficient to match the expressions for $V(k)$ at $m^2=0$.

    As before,  
\beq
    V(k) = - {1\over k^2}{1 \over {1\over g^2 N} - (J_1 - J_2)}
\label{VJJ}
\eeq
Starting with dimensional regularization and taking $m^2=0$, we have for $J_1$ 
\beq
J_1 = {1 \over 48 \pi^2}\left(12({1\over \e} - \g + \log 4\pi - \log {k^2\over \m^2}) + 28 - 24\log 2\right) \non \\
\eeq
while for $J_2$, defining $n=3-2\e$,
\beq 
    J_2&=&\pi (n-1){\m^{2\e} \over k^2} \int\frac{d^np}{(2\pi)^{n+1}} \frac{\omega_p-\omega_{p-k}}{\omega_{p-k}(\omega_p+\omega_{p-k})}
\non \\
& & \qquad \times \left[ 1-\frac{1}{n-1}\ \frac{p^2k^2-(\vec{p}\cdot\vec{k})^2}{p^2(\vec{p}-\vec{k})^2}\right]\ .
\eeq
The second term in the squared parenthesis containing the combination $\frac{p^2k^2-(\vec{p}\cdot\vec{k})}{p^2(\vec{p}-\vec{k})^2}$ leads to a convergent integral which can be done directly at $n=3$ with the result
\begin{equation}\label{IB}
    \frac{1}{48 \pi^2}\ \left(16 -24 \log2\right)\ .
\end{equation}
For the first term containing the unity, it is better to go back to $D=4-2\e$ dimensions using the identity 
\begin{equation} 
    \int_{-\infty}^{\infty} dp_4\ \frac{\vec{p}^2-p_4^2}{\left(p_4^2+\omega_p^2\right)\left(p_4^2+\omega_{p-k}^2\right)}=\pi\ \frac{\omega_p-\omega_{p-k}}{\omega_{p-k}(\omega_p+\omega_{p-k})} \ .
\label{Id}
\end{equation}
One then obtains
\begin{equation}\label{IA}
    \frac{1}{48 \pi^2}\ \left((\frac{1}{\epsilon}-\gamma+\log4\pi)-\log{k^2\over \m^2}+\frac{5}{3}\right)\ .
\end{equation}
Adding the two contributions, we find
 \begin{equation}\label{J2dimreg}
     J_2=\frac{1}{48 \pi^2}\ \left(\ (\frac{1}{\epsilon}-\gamma+\log4\pi-\log {k^2\over \m^2})+\frac{53}{3}-24 \log2 \right)\ ,
\end{equation}
and, altogether
\begin{equation}\label{finaldimreg}
    J_1-J_2=\frac{1}{48 \pi^2}\ \left(11\ (\frac{1}{\epsilon}-\gamma+\log4\pi-\log {k^2\over \m^2})+\frac{31}{3}\right)\ .
\end{equation} 
When we compute the potential $V(k)$ with a renormalized coupling $g^2_{\mathrm{\overline{MS}}}(\m)$ in the 
$\overline{MS}$ scheme, then the terms proportional to $\frac{1}{\epsilon}-\gamma+\log4\pi$ can be dropped.

    Next we turn to the cutoff regulator.  In this $m^2=0$ case 
\begin{eqnarray}\label{J13CO} 
   J_1&=&\frac{3}{2}  \int \frac{d^3p}{(2\pi)^3}\ \frac{1-u^2}{p (p^2+k^2-2 p k u)}\\
   &=& \frac{1}{48\pi^2} \ \left( 12 \ \log\frac{\Lambda^2}{k^2}+8\right) \ ,
\end{eqnarray}
where $\Lambda$ is the momentum cutoff in this integral. 
For $J_2$, the answer in cutoff regularization is
\beq\label{J2CO} 
 J_2 &=& {1\over k^2} \int\frac{d^3p}{(2\pi)^3}\  \frac{\omega_p-\omega_{p-k}}{\omega_{p-k}(\omega_p+\omega_{p-k})}\left[ 1-\frac{1}{2}\frac{p^2k^2-(\vec{p}\cdot\vec{k})^2}{p^2(\vec{p}-\vec{k})^2} \right] \non \\
\non \\
   &=&\frac{1}{48 \pi^2}\ \left(\log\frac{\Lambda^2}{k^2}+14-22 \log2\right)\ .
\eeq
Consequently, the final result is
\begin{equation}\label{finalcutoff}
    J_1-J_2=\frac{1}{48 \pi^2}\ \left(11 \log\frac{\Lambda^2}{k^2}-6+22 \log2\right)\ .
\end{equation}
    
    Now we equate the Coulomb potentials computed with momentum cutoff and dimensional regularization at $m^2=0$. This means equating the denominators of \rf{VJJ}, which is just the matching condition \rf{matching}:
\begin{eqnarray}\label{match} 
\lefteqn{\frac{1}{g^2(\Lambda)N} - \frac{1}{48 \pi^2} \left(11 \log\frac{\Lambda^2}{k^2}-6+22 \log2\right)}
\non \\
% &=& 
%     \frac{1}{(g^2_{\mathrm{\overline{MS}}})_{\mathrm{Bare}}\, N}-\frac{1}{48 \pi^2}\ \left(11\ (\frac{1}{\epsilon}-\gamma+%\log4\pi-\log k^2)+\frac{31}{3}\right)  \non \\
  & &  \qquad =  \frac{1}{g^2_{\mathrm{\overline{MS}}}(\mu)N} -\frac{1}{48 \pi^2}\ \left(11 \log\frac{\m^2}{k^2}+\frac{31}{3}\right) \ . \\\non
\end{eqnarray}
Therefore $g^2(\Lambda)N$ defined by
\begin{equation}\label{match2}
    \frac{1}{g^2(\Lambda)N}= \frac{1}{g^2_{\mathrm{\overline{MS}}}(\mu)N}-\frac{1}{48 \pi^2}\ \left(11 \log\frac{\mu^2}{\Lambda^2}+\frac{49}{3}-22\log2\right) \; ,
\end{equation}
in terms of the $\mathrm{\overline{MS}}$ running coupling $g^2_{\mathrm{\overline{MS}}}(\mu)$, is the coupling to be used to convert the momentum cutoff result to $\MS$. Defining $\Lambda_\mathrm{\overline{MS}}$ by the equation
\begin{equation}\label{Lambda}
    \frac{1}{g^2_{\mathrm{\overline{MS}}}(\mu)N}= \frac{11}{48 \pi^2} \ \log\frac{\mu^2}{\Lambda^2_\mathrm{\overline{MS}}}\ ,
\end{equation}
one finds that
\begin{equation}\label{finalmatch}
     \frac{1}{g^2(\Lambda)N}=\frac{1}{48 \pi^2}\ \left(11 \log\frac{\Lambda^2}{\Lambda^2_\mathrm{\overline{MS}}}-\frac{49}{3}+22\log2\right)
\end{equation}
is the choice of $g^2(\L)$ required to convert our result from cutoff regularization to the $\MS$ scheme.

    A useful check of this method for converting cutoff regularization to the $\MS$ scheme is to go back to the massive propagator case in section \ref{massive_case}, recalculate $J_1$ and $J_2$ with the cutoff regulator, and insert those values plus \rf{finalmatch}  into \rf{VJJ}.  When this is done, we find that our result agrees precisely with the result already obtained using dimensional regularization and $\MS$ subtraction, shown in eq.\ \rf{Vmassive}. \\

\section{The Coulomb potential, ~final result}

    Inserting \rf{finalmatch} into \rf{Vinv}, we see that $\L^2$ cancels out, and the potential, in terms of the physical
scale $\LM$, is
\begin{widetext}
\beq
\label{Coulomb} 
 V(k) =  - {1\over k^2}\left[ \frac{11}{48 \pi^2}\left(\log{2 m^2\over \LM^2}-\frac{59}{33}\right)\non  + \frac{1}{48\pi^2}\frac{k^2}{m^2}\left( \frac{41}{20}\log\frac{m^2}{k^2}-\frac{\log2}{4}+\frac{219}{25} \right) \right]^{-1} \ . \non \\
\eeq
\end{widetext}
As in the case of the massive transverse propagator, we now set $m^2$ to the unique value at which power behavior of the Coulomb potential is enhanced in the infrared.  This leads us to
\begin{equation}\label{mlambda}
    m=\frac{1}{\sqrt{2}}\Lambda_\mathrm{\overline{MS}}\ \mathrm{e}^{59/66}\approx 1.73\ \Lambda_\mathrm{\overline{MS}}\ ,
\end{equation}
and therefore, for $k^2 \ll m^2$

\begin{equation}\label{Result}
    V(k) =  -\frac{48\pi^2 m^2}{k^4\left( \frac{41}{20}\log\frac{m^2}{k^2}-\frac{\log2}{4}+\frac{219}{25} \right)}  \ ,
\end{equation}
where $m$ is given in (\ref{mlambda}). 
We have finally ended up with a potential which behaves, in the infrared, as $-1/k^4$ modified by a logarithm. 
       
       One often hears that a $-1/k^4$ potential in momentum space corresponds, upon Fourier transformation, to a linearly increasing potential in position space.  Strictly speaking, this is untrue; the Fourier transform of $-1/k^4$ is actually minus infinity, due to the very singular behavior of $1/k^4$ as $k\ra 0$.  But this is precisely why it is important to include the quark-antiquark self-energies, as we have done in eq.\ \rf{VC}.  The Coulomb self-energies of quarks and antiquarks are also infinite, and this is not only the usual UV divergence which can be regulated with,
e.g., a lattice cutoff.  The Coulomb self-energies of quarks and antiquarks have, in addition, an infrared divergence, and a short-distance or high-momentum or lattice cutoff will not make this type of self-energy finite.  In fact, this is already a reason why isolated quarks and antiquarks, or a non-singlet quark-antiquark pair, are infinitely massive, and cannot appear as asymptotic states.  But for a color singlet 
quark-antiquark pair, the infrared infinities of the self-energy and interaction terms precisely cancel, leaving only UV divergent contributions to the self-energies, and a finite interaction term.  This cancellation has been noted previously in ref.\ \cite{Greensite:2004ke}, in connection with the instantaneous Coulomb interaction, where it was shown more generally that the cancellation of infinities is exact for any global color singlet combination of static quarks and antiquarks.
  
\begin{figure}[t!]
\centerline{\scalebox{0.80}{\includegraphics{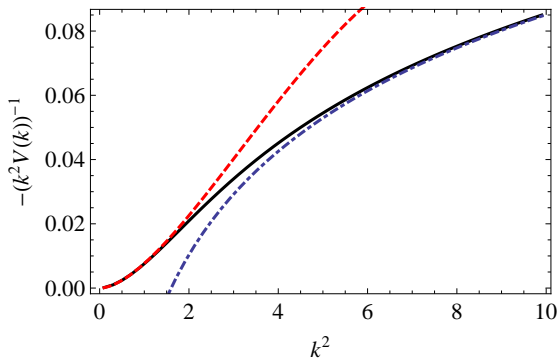}}}
\caption{Numerical calculation of $-(k^2 V(k))^{-1}$ (solid line) compared to the infrared limit (upper dashed line) derived here, and the standard one-loop perturbative result (lower dot-dash line).  The $x$-axis is in units of $m^2$.}
\label{fig1}
\end{figure}

     The color Coulomb potential is 
\beq     
         V_C(R) = - {C_r \over N} \int {d^3 k\over (2\pi)^3}  V(k) (1 - e^{i \vk \cdot \vec{R}}) 
\label{cpot}
\eeq
and, using the small-$k$ approximation \rf{Result} to $V(k)$,  
the Fourier transform to position space gives us asymptotically
\beq
         V_C(R) \stackrel{R\ra \infty}{=} {C_r \over N}  \left( {120 \pi \over 41} {m^2 \over  \log (8.12 m R/3)^2}\right) R \ .
\eeq
This transform is carried out in Appendix \ref{appB}.  However, the 
small-$k$ approximation is only valid at large distances, i.e.\ $R \gg 1/m$, in which case the integral is sensitive mainly to the small $k$ behavior of $V(k)$.   An expression for $V(k)$ valid at all $k$ will agree with \rf{Result} at small $k$, and the usual perturbative result \rf{Vpt} at large $k$.  We do not have an analytical expression for $V(k)$ valid at all $k$, but it is not hard to compute $V(k)$ numerically, by evaluating $J_1$ and $J_2$ in eqs.\ (\ref{J1}-\ref{J2}) numerically.
The result for $(-k^2 V(k))^{-1}$ in cutoff regularization is shown in Fig.\ \ref{fig1}, and it interpolates nicely between our analytical result at small $k$, and the perturbative result for large $k$ at $m^2=0$.

\begin{figure}[t!]
\centerline{\scalebox{0.80}{\includegraphics{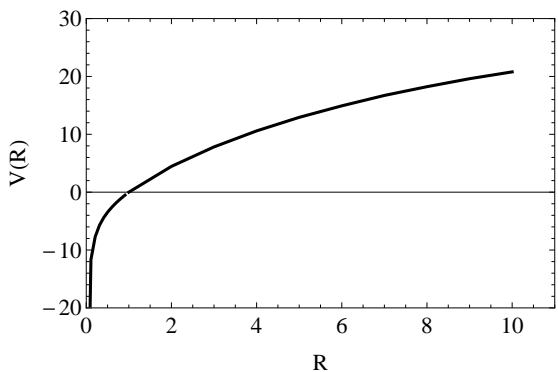}}}
\caption{Potential $V(R)$ vs.\ $R$, obtained from a Fourier transform to position space of the numerical solution for 
$V(k)$.  The result includes the self-energy, which is both ultraviolet and infrared divergent. The infrared divergence 
is cancelled, for a color singlet, by a corresponding term in the interaction, as explained in the text.  To regulate the ultraviolet divergence  we have made an arbitrary subtraction such that $V(R)$ passes through zero at $R=1$.   
$V(R)$ is in units of $m$, $R$ in units of $1/m$.}
\label{fig2}
\end{figure}

     The next step is to Fourier transform our result for $V(k)$ to a potential in position space, i.e.\ eq.\ \rf{cpot}.
Using the numerical result for $V(k)$ at all momentum, of course the UV divergence of the self-energy will appear.
On the lattice this UV divergence is regulated by the lattice spacing, and in the ordinary one-loop perturbative calculation of the Coulomb potential, the static quark self-energy is dropped altogether.  In our case it is simplest to get rid of the UV self-energy divergence by  making an arbitrary subtraction, such that the potential vanishes at $R=1$ (in units of $1/m$); i.e.\ we compute $V(R)-V(1)$.  The result is shown in Fig.\ 2.

     It is interesting to compare our result with lattice data at large-$N$.  Of course one cannot directly compare string tensions, because of logarithmic modification of the linear term.  The best one can do is to compare the slope of
$V(R)$ in Fig.\ 2,  multiplied by the large-$N$ Casimir factor $C_F/N=\oh$,  
with the lattice result for the asymptotic string tension, extrapolated to large-$N$.  In order to make this comparison, we need the expression for the lattice asymptotic string tension in units of $\LM$.  This has been derived in ref.\ \cite{Allton:2008ty} (see also \cite{Lucini:2008vi}), which finds that $\LM/\sqrt{\s}=0.503(2)(40)$ at $N\ra \infty$, where the uncertainties refer to statistical error, and an estimate of the systematic error from all sources.  Therefore the string tension at large-$N$, derived from lattice Monte Carlo simulations, is $\s = 3.95 \LM^2$.  
Since $V_C(R) = \oh V(R)$ doesn't really have an asymptotic string tension, the comparison with $\s$ depends on where we choose compute the slope of the $V_C(R)$.   At, e.g., $R=2m^{-1} \approx 1.16 \LM^{-1}$, where a confining potential seems to have taken over from $1/R$ behavior, we find~\footnote{For finite $N$, $\s_{coul}$ at $R=2m^{-1}$ would be
$5.69(1-1/N^2)\LM^2$.} 
\beq
\s_{coul_{|_{R=2/m}}} &\equiv& \left({dV_C\over dR}\right)_{|_{R=2/m}} \approx 1.9 m^2
\non \\
&\approx& 5.69 \LM^2
\non \\
 &\approx& 1.44 \s \ .
\eeq

\section{Conclusions}

    In this article we have explored the idea that, in Coulomb gauge, restriction to the Gribov region can be approximated by a momentum-dependent mass term in the action.  Within the Gribov region, the bulk of configurations
should lie near the horizon, and configurations near the horizon are expected to strengthen the
long-range behavior of the color Coulomb potential.   If the mass term should have this same effect, by suppressing (on average) configurations outside the Gribov horizon, then the mass parameter should be adjusted to the unique value at which the Coulomb potential is enhanced in the infrared.

    We have tested this idea at the one-loop level, by a perturbative calculation of the non-instantaneous color Coulomb potential derived from $g^2 D_{44}(\vk,k_4=0)$.  For a momentum-independent mass term, the finding is that the     
infrared behavior is confining, but only marginally; the potential rises logarithmically with quark-antiquark separation.
However, for a momentum-dependent mass term leading to the propagator suggested by Gribov, the result is quite different:  we find a confining potential rising as linear modified by a logarithm, and our potential is expressed in terms
of the usual scale $\LM$.  

    This result, like most of its kind, must be interpreted with caution.  In the first place, we have no idea how accurate our one-loop result may be.  The best check would be to carry out the calculation further, to two loops, but this is a formidable task in Coulomb gauge.  In the second place, we cannot be sure of the validity of the Gribov propagator in Coulomb gauge.  At present the lattice Monte Carlo evidence is suggestive but not decisive on this point \cite{Burgio:2008jr,Nakagawa:2011ar}, and we hope that our work will help to motivate further lattice investigations of this issue.
Finally, because of the logarithmic modification, the potential found here is certainly not an upper bound on the static quark potential.  However, the upper bound derived by Zwanziger \cite{Zwanziger:2002sh} only applies to the instantaneous Coulomb potential, rather than the full one-gluon exchange potential.  It would be interesting to derive the instantaneous potential at one loop, along the lines we have followed here.  Such a potential would be of particular interest for variational calculations of bound states and the gluon chain, and for this purpose the validity of the potential in an intermediate range of distances may be sufficient.  We leave this case for future investigation.

\acknowledgments{JG and MG acknowledge support by
the U.S.\ Department of Energy under Grant No.\ DE-FG03-92ER40711. SP is supported by FPA2008-01430, FPA2011-25948, SGR2005-00916, the Spanish Consolider-Ingenio 2010 Program CPAN (CSD2007-00042) and by  the Programa de Movilidad PR2010-0284. APS 
 acknowledges support by
the U.S.\ Department of Energy under Grant No.\ DE-FG02-87ER40365.   

\appendix

\section{Application of the Mellin-Barnes transform} 

    The Mellin-Barnes transform $M(f,s)$ of a function $f$, is defined by
\beq
     M(f,s) &=& \int_0^\infty dx ~ x^{s-1} f(x) \ ,
\eeq
and the corresponding inverse transformation is
\beq
        f(x) = {1\over 2\pi i} \int_\Gamma ds ~ x^{-s} M(f,s) \ .
\label{Aa}
\eeq
The first integral is typically well-defined in a region (the ``fundamental strip") of the complex $s$-plane, with 
${s_{min}< Re(s) <s_{max}}$, and the contour $\Gamma$ is a line parallel to the imaginary axis inside the fundamental strip.   Now let
\beq
M(f,s) \asymp \sum_{p,k} {r_{pk} \over (s+p)^k}
\eeq
be a ``singular expansion" (denoted ``$\asymp$") of $M(f,s)$ on the left-hand side of the fundamental strip. A singular expansion is obtained by keeping all the singular terms in the Laurent series around each pole of $M(f,s)$, in this case restricted to poles on the left-hand side of the fundamental strip.  Then the Converse Mapping Theorem tells us that asymptotically, as $x \ra 0$,
\beq
            f(x)  \sim \sum_{p,k} {(-1)^{k-1} \over (k-1)!} r_{pk} x^p \log^{k-1}x \ .
\eeq
A proof of the converse mapping theorem is given in \cite{Flajolet19953}, and application to Feynman diagrams
is found in \cite{Friot:2005cu}.

The strategy is to put the integrals in eqs.\ \rf{bad1}, \rf{bad2} in the form \rf{Aa}, make a singular expansion of $M(f,s)$, and apply the converse mapping theorm.  For this purpose, we will need the Mellin-Barnes representation \cite{WhWa27}
\beq
{1 \over (1+A)^\n} = {1\over 2\pi i} \int_\G ds ~ A^{-s} {\G(s) \G(\n-s) \over \G(\n)} \ .
\eeq
The fundamental strip is in the region $0<Re(s)<\n$.

    Beginning with the integral \rf{bad2}, we apply the above identity with $\n=1$
\beq
I &=& k^2 \int dx_1 dx_2 {\theta(1-x_1-x_2)  x_2^2 \over m^2 x_1^{3/2}} 
\non \\
  & &   \times {1\over 2\pi i} \int_\G ds \left({k^2 x_2(1-x_2) \over m^2 x_1}\right)^{-s}  {\pi \over \sin(\pi s)} \ .
\eeq
Interchanging orders of integration, the integrals over $x_1,x_2$ can be carried out exactly, with the result
\beq
I = {1 \over 2 \pi i} {k^2 \over m^2} \int_\G ds ~ \left({k^2 \over m^2}\right)^{-s} \left[  {\pi \over \sin(\pi s)}
     {2 \sqrt{\pi} \G(3-s) \over (2s-1) \G({7\over 2} -s)} \right] \ .
\non \\
\eeq
Now making a singular expansion, and applying the converse mapping theorem, we have
\beq
I &=&   {1 \over 2 \pi i} {k^2 \over m^2} \int_\G ds ~ \left({k^2 \over m^2}\right)^{-s} \left[ {3\pi^2/8 \over s-\oh}
 - {32 \over 15 s} + {64 \over 105(s+1)}
  + ... \right]
\non \\
&=& {3 \pi^2 \over 8} \left({k^2 \over m^2}\right)^\oh - {32\over 15} {k^2 \over m^2} + ... \ .
\eeq

    The integral in \rf{bad1} is handled in a similar way.  First write
\beq
I' &=& \int dx_1 dx_2 \theta(1-x_1-x_2) x_1^{-\oh} 
\non \\
   & & \times \log\left[{m^2 x_1 \over \LM^2}
      \left(1 + {k^2 x_2(1-x_2) \over m^2 x_1}\right)\right] \ ,
\eeq
and use the identity
\beq
\log(1+A) = {1 \over 2\pi i} \int_\G ds ~ A^{-s}  {\pi/s \over \sin(\pi s)} \ ,
\eeq
where the fundamental strip is in the region $-1< Re(s) < 0$.  Then
\beq
I' &=& \int_0^1 dx_2 \int_0^{1-x_2} dx_1 ~ x_1^{-1/2}\left( \log{m^2 \over \LM^2} + \log x_1 \right)
\non \\
& & + \int_0^1 dx_2 \int_0^{1-x_2} {dx_1 \over \sqrt{x_1} }{1\over 2\pi i} \int_\G ds \left({k^2 x_2(1-x_2) \over m^2 x_1}\right)^{-s} {\pi/s \over \sin\pi s} \ . \non \\
\eeq
Again interchanging orders of integration, carrying out the integrations over $x_1,x_2$, and making a singular expansion,
we find
\beq
I' &=& {4 \over 3} \log{m^2 \over \LM^2} - {32 \over 9} + {1\over 2\pi i} \int_\G ds  \left({k^2\over m^2}\right)^{-s}
\non  \\
& & \qquad \times {\pi/s \over \sin\pi s}  
    { \sqrt{\pi} \G(1-s) \over (1+2s) \G({5\over 2} -s)}
\non \\
&=&  {4 \over 3} \log{m^2 \over \LM^2} - {32 \over 9} + {1\over 2\pi i} \int_\G ds  \left({k^2\over m^2}\right)^{-s}
\non \\
& & \qquad \times    \left( {\pi^2/4 \over s+\oh} - {8/15 \over s+1} + ... \right)
\non \\
&=& {4 \over 3} \log{m^2 \over \LM^2} - {32 \over 9} + {\pi^2 \over 4} \left({k^2 \over m^2}\right)^\oh 
- {8\over 15} {k^2 \over m^2} + ... \ .
\non \\
\eeq
This completes the low-$k^2$ evaluation of the integrals in (\ref{bad1}), (\ref{bad2}).

\section{Transform to position space \label{appB}}
 
In order to determine the asymptotic form of the one-loop potential as $R\to \infty$, we will need to transform the momentum-space expression at small $k$, eq.\ \rf{Result}, to position space.  Absorbing a constant $-\oq \log 2
+ {219\over 25}$ into the logarithm, $V(k)$ can be written 
\beq
V(k)  &=& {960 \pi^2 m^2 \over 41} \frac{1}{k^4 \log \frac{k^2}{(8.12m)^2}}  \label{v}
\non \\
&=& {960 \pi^2 m^2 \over {41}} \V(k)  \ .
\label{VkB}
\eeq
This is expected to yield a positive linear potential, modulo logarithms, plus an infinite constant which is removed by the self-energy term, as discussed earlier.  

   Before proceeding, we should stress again that \rf{VkB} is only valid at small
$k^2 \ll m^2$.  The excuse for taking the Fourier transform anyway is that the large-$R$ behavior we are interested in is dominated by small $k$ behavior, so the error at large $k$ should only affect terms which are subleading in $R$.  
Note in particular that there is an unphysical Landau pole in \rf{VkB} on the real axis, at a comparatively high momentum 
$k=8.12m=14 \LM$.  This pole is certainly not present in the result we have obtained numerically for $V(k)$ at all momenta, which is  displayed in Fig.\ \ref{fig1}.  The Fourier transform of \rf{VkB} will nonetheless require a prescription (e.g.\ principal value) for dealing with the unphysical pole, but the choice of prescription, as we will see, only introduces an ambiguity in subleading terms at large $R$.

In the following we will switch to units $\widetilde{m}=8.12m=1$, so that
\beq
\V(k) = \frac{1}{k^4\log k^2}  \  .
\eeq
The inverse log has a cut on the negative axis and a Landau pole at $k^2 = 1$. The discontinuity across the cut is easily evaluated, and the $1/\log k^2$ factor can be expressed through a Cauchy integral 
\begin{equation} 
\frac{1}{\log k^2} = \int_0^\infty ds \frac{\rho(s)}{s + k^2}   + \frac{1}{k^2 - 1},\; \mbox{ with }  \rho(s) = \frac{1}{\log^2 s + \pi^2} \ . \label{c} 
\end{equation} 
In order to perform the Fourier transform, the IR singularity of the $1/k^4$ term is regularized by writing 
\begin{equation}
\frac{1}{k^4} \to \lim_{\mu \to 0} \frac{1}{k^2 (k^2 + \mu^2)} \ ,
\end{equation} 
whose Fourier transform leads to an additional  constant (infinite in the $\mu \to 0$ limit), which is removed by the self-energy
term.  This removal amounts to subtracting $V(0)$ from $V(R)$.

In the following we consider the dispersive (first term in the r.h.s.\ of \rf{c})  and Landau pole contributions separately 
\beq
\widetilde \V(R) &=& \V(R) - \V(0) 
\non \\
&=& [V_D(R) - V_D(0) ]   + [V_P(R) - V_P(0)] \ ,
\eeq
where
\beq
\lefteqn{V_D(R) - V_D(0)}
\non \\
&=&  \int \frac{d^3k}{(2\pi)^3} [e^{i \vk \cdot \vR} - 1] V_D(k) 
\non \\  
&=& \lim_{\mu \to 0}  \int \frac{d^3k}{(2\pi)^3} [e^{i \vk \cdot \vR} - 1] 
\frac{1}{k^2(k^2 + \mu^2)} \left[ \frac{1}{\log k^2} - \frac{1}{k^2 - 1} \right] \ ,
\non \\
\eeq 
and
\beq 
 \lefteqn{V_P(R) - V_P(0)}
\non \\
  &=&  \int \frac{d^3k}{(2\pi)^3} [e^{i \vk \cdot \vR} - 1] V_P(k) 
\non \\
&=& \lim_{\mu \to 0}  \int \frac{d^3k}{(2\pi)^3} [e^{i \vk \cdot \vR} - 1] 
\frac{1}{k^2(k^2 + \mu^2)}  \frac{1}{k^2 - 1} \ .
\non \\
\eeq

For the dispersive part one finds 
\beq
\lefteqn{\widetilde \V(R)}
\non \\
&=& \lim_{\mu \to 0} \int \frac{d^3k}{(2\pi)^3}  \int_0^\infty ds \rho(s) 
[e^{i \vk \cdot \vR} - 1]
\frac{1}{\mu^2} \left( \frac{1}{k^2} - \frac{1}{k^2 + \mu^2}\right) \frac{1}{k^2 + s}  \nonumber \\
&=&  
\lim_{\mu \to 0} \int \frac{d^3k}{(2\pi)^3}  \int_0^\infty ds \rho(s) 
[e^{i \vk \cdot \vR} - 1] \frac{1}{\mu^2} 
\non \\
& & \times \left( \frac{1}{s} \frac{1}{k^2} - \frac{1}{s} \frac{1}{k^2 + s} 
 + \frac{1}{\mu^2 - s} \frac{1}{k^2 + \mu^2} - \frac{1}{\mu^2 - s} \frac{1}{k^2 + s} \right) 
\nonumber \\
 &=& \lim_{\mu \to 0} \int_0^\infty ds \rho(s) \frac{1}{\mu^2} \bigg\{ \frac{1}{4\pi R} \bigg[ \frac{1}{s}[1 - e^{-R\sqrt{s}}] 
 \non \\
  & & \qquad   + \frac{1}{\mu^2 -s } [e^{-R \mu } - e^{-R \sqrt{s}} ] \bigg] - (R \to 0) \bigg\} 
\nonumber \\
 &=&  -\frac{R}{8\pi}   \int_0^\infty \frac{ds}{s}  \rho(s) \left[  1 + 2\frac{1 - e^{-R\sqrt{s}}}{ s R^2} 
  - \frac{2}{R \sqrt{s}} \right]  \ .
\end{eqnarray} 
Change variables $s \to z = s R^2$,
\beq
\lefteqn{V_D(R) - V_D(0)}
\non \\
  &=& -\frac{ R}{8\pi} \int_0^\infty \frac{dz}{z} \frac{1}{\log^2\frac{z}{R^2} + \pi^2}
  \left[ 1 + 2 \frac{1 - e^{-\sqrt{z}}}{z} - \frac{2}{\sqrt{z}} \right] \ .
  \non \\
\eeq
Next, the term in the bracket is approximated by 
 \begin{equation} 
 1 + 2\frac{1 - e^{-\sqrt{z}}}{z} - \frac{2}{\sqrt{z}} \to \frac{\sqrt{z}}{\sqrt{z} + 3} \label{r} \ ,
 \end{equation}
which has the same limit in for both $z \to 0$ and $z \to \infty$, limits.  Therefore in the integral 
\begin{equation} 
  \int_0^\infty \frac{dz}{z} \frac{1}{\log^2\frac{z}{R^2} + \pi^2} \left[ \left( 1 + 2 \frac{1 - e^{-\sqrt{z}}}{z} - \frac{2}{\sqrt{z}} \right)  -  \frac{\sqrt{z}}{\sqrt{z} + 3}   \right]
\end{equation} 
one can take the $R \to \infty$ limit since the resulting integral is convergent. This gives a contribution of  
$O(R/\log^2R)$, which, as will be shown below, is subleading  in the 
$R \to \infty$ limit, since the leading behavior is $O(R/\log R)$.

In the $R \to \infty$ limit, the leading behavior can therefore be obtained from 
 \begin{equation}
V_D(R) - V_D(0)  \approx  -\frac{ R}{8\pi} \int_0^\infty \frac{dz}{z} \frac{1}{\log^2\frac{z}{R^2} + \pi^2}  
 \frac{\sqrt{z}}{\sqrt{z} + 3} \ ,
\end{equation} 
which after a few more manipulations can be written as 
\begin{eqnarray}
\lefteqn{V_D(R) - V_D(0)}
\non \\
  &\approx&  -\frac{ R}{8\pi} \int_0^\infty \frac{dz}{z} \frac{1}{\log^2\frac{z}{R^2} + \pi^2}  
 \frac{\sqrt{z}}{\sqrt{z} + 3}  
\non \\
&=&   -\frac{ R}{8\pi} \int_0^\infty \frac{dz}{z} \frac{1}{\log^2\left(z\left({3\over R}\right)^2\right)  + \pi^2}  
 \frac{\sqrt{z}}{\sqrt{z} + 1}    
 \nonumber \\
 &=&  -\frac{ R}{8\pi} \int_0^\infty \frac{dz}{z} \frac{1}{\log^2\left(z\left({3\over R}\right)^2\right)  + \pi^2}   
 \non \\
 & & \qquad   +\frac{ R}{8\pi} \int_0^\infty \frac{dz}{z} \frac{1}{\log^2\left(z\left({3\over R}\right)^2\right)  + \pi^2}  
 \frac{1}{\sqrt{z} + 1}  
 \nonumber \\
 &=&   -\frac{ R}{8\pi}    +\frac{ R}{8\pi} \int_0^1  \frac{dz}{z} \frac{1}{\log^2\left(z\left({3\over R}\right)^2\right)  + \pi^2}  
 \frac{1}{\sqrt{z} + 1} 
 \non \\
& & \qquad +\frac{ R}{8\pi} \int_1^\infty \frac{dz}{z} \frac{1}{\log^2\left(z\left({3\over R}\right)^2\right)  + \pi^2}  
 \frac{1}{\sqrt{z} + 1}  \ .
  \nonumber \\ \label{dis} 
 \end{eqnarray} 
The last integral is finite in the limit $R \to \infty$, again leading to a term of the order of $O(R/\log^2 R)$. The remaining integral is dominated by $z = 0$ and  $1/(\sqrt z+1)$ can be expanded in powers of $\sqrt{z}$ leading to, in the limit $R \to \infty$, 
\beq 
 \lefteqn{ \frac{ R}{8\pi} \int_0^1  \frac{dz}{z} \frac{1}{\log^2\left(z\left({3\over R}\right)^2\right)  + \pi^2}   \frac{1}{\sqrt{z} + 1} }
\non \\
  & & \qquad \qquad  = \frac{R}{8\pi \log (R/3)^2} 
 + O\left( \frac{R}{\log^2 R}\right) \ .
 \eeq

We now return to the pole term.   For this we need to evaluate
\begin{equation}
\int_P\ \frac{d^3k}{(2\pi)^3}\,e^{i{\vec k}\cdot{\vec R}}\frac{1}{k^2-1}\ .
\end{equation}
This integral is not well-defined, because there is a pole at $k=1$ (or, in general units, $k=8.12m$)  on the positive real axis.  The leading $R$ dependence, however, does not depend on how 
 the pole is circumvented. This is because, in the  neighborhood of the pole, $k$ is finite, while 
  the leading-$R$ behavior is determined by the behavior of the integrand in the $k \to 0$ limit. 
   To illustrate this point, we consider a prescription ``P''  for how to skip the pole which excludes 
    from the integration range the interval $1-b \varepsilon\leq k \leq  1+ a\varepsilon$, 
    \begin{equation}
\int_P\frac{d^3k}{(2\pi)^3}\,e^{i{\vec k}\cdot{\vec R}}\frac{1}{k^2-1}
=\frac{\cos{(R)}+ \alpha \sin{(R)}}{4\pi R}\ .
\end{equation}
where $\alpha=\log(a/b)$ parametrizes the ambiguity. For instance, for the principal-value prescription we have that $a=b$, and thus $\alpha=0$.
For arbitrary $\alpha$, one finds
\begin{eqnarray}
\lefteqn{V_P(R) - V_P(0)}
\nonumber \\
& & \qquad=  \int \frac{d^3k}{(2\pi)^3} [e^{i {\vec k} \cdot {\vec R}} - 1] V_P(k)
\nonumber \\
& & \qquad = \lim_{\mu \to 0}  \int \frac{d^3k}{(2\pi)^3} [e^{i {\vec k} \cdot {\vec R}} - 1]
\frac{1}{k^2(k^2 + \mu^2)}  \frac{1}{k^2 - 1}
\nonumber \\
& & \qquad =  \frac{R}{8\pi} \left( 1 - \frac{2\alpha}{R}
- 2\frac{1 - (\cos R+\alpha\sin R)}{R^2} \right)  = \frac{R}{8\pi} + O(1) \ .
\non \\
 \label{alphaB}
 \end{eqnarray}
%As advertised, the ambiguity in prescription does not at all affect asymptotic behavior, but only %terms which are non-leading at large $R$. 
%   We now return to the pole term. The real part of FT of $1/(k^2 - m^2)$ with $m$ real is given 
%   by~\footnote{ The integral is defined trough analytical continuation in $z = -m^2$, which leads to 
%    an  analytical function in the entire $z$-plane cut along the negative-$z$ axis. The  real part in  Eq.~(\ref{rft}) is given by the boundary value on the cut.} 
%\begin{equation}
%\int \frac{d^3k}{(2\pi)^3} e^{i \vk \cdot \vR} \frac{1}{k^2 - m^2} = \frac{\cos(mR)}{4\pi R} \ , \label{rft} 
%\end{equation}
%which for the pole term gives  
%\begin{eqnarray} 
%\lefteqn{V_P(R) - V_P(0)}
%\non \\
%& & \qquad=  \int \frac{d^3k}{(2\pi)^3} [e^{i \vk \cdot \vR} - 1] V_P(k) 
%\non \\
%& & \qquad = \lim_{\mu \to 0}  \int \frac{d^3k}{(2\pi)^3} [e^{i \vk \cdot \vR} - 1] 
%\frac{1}{k^2(k^2 + \mu^2)}  \frac{1}{k^2 - 1} 
%\nonumber \\ 
%& & \qquad =  \frac{R}{8\pi} \left( 1 - 2\frac{1 - \cos R}{R^2} \right) \ .
 %\end{eqnarray}
 In the limit of $R \to \infty$, \rf{alphaB} reduces to  $R/8\pi$ which cancels the corresponding term in dispersive part, cf.\ \rf{dis}. So finally the leading behavior in the large-$R$ limit is given by 
 \beq
 \widetilde \V(R) &\stackrel{R \to \infty}{=}&  \frac{R}{8\pi \log (R/3)^2}  + O\left( \frac{R}{\log^2 R}\right) \ ,
 \eeq
 or asymptotically, restoring constants and factors of $m$
 \beq
 V(R) \sim \left( {120 \pi \over 41} {m^2 \over  \log (8.12 m R/3)^2}\right) R \ .
 \eeq

\bibliography{coul}

\end{document}